# ANALYSIS AND PACKAGING OF RADIOCHEMICAL SOLAR NEUTRINO DATA: A BAYESIAN APPROACH

P.A. Sturrock[1], and M.S. Wheatland[2]

[1]Center for Space Science and Astrophysics, Stanford University, Stanford, CA 94305-4060
(email: sturrock@stanford.edu)
[2]School of Physics, University of Sydney, Sydney, NSW 2006, Australia
(email: wheat@physics.usyd.edu.au)

## ABSTRACT

According to current practice, the results of each run of a radiochemical solar neutrino experiment comprise an estimate of the flux and upper and lower error estimates. These estimates are derived by a maximum-likelihood procedure from the times of decay events in the analysis chamber. This procedure has the following shortcomings: (a) Published results sometimes include negative flux estimates. (b) Even if the flux estimate is non-negative, the probability distribution function implied by the flux and error estimates will extend into negative territory; and (c) The overall flux estimate derived from the results of a sequence of runs may differ substantially from an estimate made by a "global" analysis of all of the timing data taken together. These defects indicate that the usual "packaging" of data in radiochemical solar neutrino experiments provides an inadequate summary of the data, which implies a loss of information. This article reviews this problem from a Bayesian perspective, and suggests an alternative scheme for the packaging of radiochemical solar neutrino data, which is we believe free from the above objections.

# 1. Introduction

One class of solar neutrino experiments (historically, the first) employs radiochemical techniques to estimate the neutrino flux. The Homestake (Cleveland et al., 1995; Cleveland et al., 1998; Davis and Cox, 1991; Lande et al., 1992), GALLEX-GNO (Altmann et al. 2000; Anselmann et al., 1992; Anselmann et al., 1993; Anselmann et al., 1995; Hampel et al., 1996; Hampel et al., 1999), and SAGE (Abdurashitov et al. 1995; Abdurashitov et al. 1999) experiments belong to this class. In such experiments, one arrives at estimates of the solar neutrino flux by a multi-stage process. (For a short summary, see for instance, Bahcall 1989.)

In the first stage, neutrinos are "captured" in a large tank containing 133 tons of perchlorethylene in the Homestake experiment; 101 tons of gallium chloride ($GaCl_3$) in the GALLEX and GNO experiments; or 50 metric tons of gallium in the SAGE experiment. In the Homestake experiment, the capture of a neutrino by a $^{37}Cl$ atom converts it into a $^{37}Ar$ atom which is radioactive, decaying back to $^{37}Cl$ with a half-life of 35.0 days. In the GALLEX, GNO, and SAGE experiments, the capture of a neutrino by a $^{71}Ga$ atom converts it into a radioactive $^{71}Ge$ atom, which decays back to $^{71}Ga$ with a half-life of 11.43 days.

In the second stage, the capture-produced atoms are extracted from the tank by chemical processes, and transferred to a small analysis chamber.

The third stage comprises the detection of decay events in the analysis chamber. When a $^{37}Ar$ atom decays back to a $^{37}Cl$ atom, the decay is accompanied by the emission of an Auger electron with energy 2.82 keV. When a $^{71}Ge$ atom decays back to a $^{71}Ga$ atom, the decay is accompanied by the emission of an Auger electron with energy 1.2 keV if the electron comes from the L shell, or 10.4 keV if it comes from the K shell. These events may be detected by proportional counters. However, there are inevitably also background events unrelated to these decay events.

The fourth stage comprises an analysis of the times of events registered by the proportional counters. This analysis is the subject of this article. Since radioactive decay is a stochastic process, it is necessary to use statistical methods to infer the neutrino flux from timing data, comprising a list of times at which decay or decay-like events are detected. Radiochemical neutrino experiments have traditionally used a maximum-likelihood procedure developed in a seminal work by Cleveland (1983), which we here refer to as the "likelihood" analysis procedure. The problem of analyzing radiochemical timing data was subsequently re-analyzed independently, using a different procedure, by Opendak and Wildenhain (1994), who arrived at the same end result as Cleveland.

The likelihood method has served well over the years, especially when one is assuming the flux to be constant. However, there appear to be some shortcomings:

(a) As used by some collaborations (Altmann et al. 2000; Anselmann et al., 1992; Anselmann et al., 1993; Anselmann et al., 1995; Hampel et al., 1996; Hampel et al., 1999), the method results in flux estimates that include negative values;

(b) Even if the flux estimate is non-negative, a double-gaussian form for the probability distribution function implied by the flux and error estimates (such as that specified below) will yield non-zero probabilities for negative flux values;

(c) Estimation of the neutrino flux (assumed constant) incorporating information from all runs is thought to require a "global" maximum likelihood analysis of all runs taken together, which can involve as many as two hundred degrees of freedom (a background rate for each of the K and L lines for each run, plus a flux value); and

(d) The global flux estimate may differ from the estimate derived by combining all run estimates.

As an example of point (d), we refer to GNO data. The GNO collaboration has published data for 58 runs in the time interval 1998.4 to 2003.3 (Altmann et al., 2005). For each run, they provide estimates of the flux and lower and upper flux estimates, as obtained by the likelihood procedure. The results of radiochemical experiments are actually given in SNU (the solar neutrino units), which is a measure of the capture rate per atom (1 $SNU = 1 \times 10^{-36}$ captures per atom per second), but we here use the term "flux" for simplicity.

As shown in the appendix, a maximum likelihood analysis of the published data (flux and error upper and lower error estimates) leads to the estimate $44.8 \pm 4.9\,\mathrm{SNU}$. On the other hand, Cattadori (2004) has carried out a global likelihood analysis, and obtains the estimate $62.9 \pm 5.9\,\mathrm{SNU}$. The difference is $18.1\,\mathrm{SNU}$, and the error estimate of the difference is $7.7\,\mathrm{SNU}$, so that the two estimates of the flux, both based on data from all runs, differ by about 2.4 sigma. On the assumption that the global likelihood estimate is correct, it follows that an estimate made by combining the data for individual runs is incorrect. This implies that there has been a loss of information in the likelihood procedure for packaging radiochemical solar neutrino data. For the above reasons, it seems prudent to re-think the present practice of presenting the results of each run in terms of just three numbers: a flux estimate, and upper and lower error estimates.

In Section 2, we comment briefly on the neutrino-capture process. In Section 3, we discuss the analysis of timing data for a single run, and in Section 4 we discuss the merging of data to provide information concerning many runs. In Section 5, we discuss ways of incorporating supplementary experimental information into the analysis. Concluding comments are made in Section 6.

## 2. Neutrino Capture

Concerning the first stage of the measurement process, we assume that the fluid has been purged of neutrino-capture products at the start time $t_S$, and that the new capture products are extracted at the end time $t_E$. If the number of "convertible" atoms in the tank ($^{37}$Cl or $^{71}$Ga) is $N_T$, the expected number of capture-produced atoms ($^{37}$Ar or $^{71}$Ge) is given by

$$Q_T = 10^{-36} N_T \int_{t_S}^{t_E} dt\, F(t)\, e^{-\lambda(t_E - t)} ,  \qquad (1)$$

where $\lambda$ is the decay coefficient (0.0198 day$^{-1}$ for the decay of $^{37}$Ar back to $^{37}$Cl, and 0.0606 day$^{-1}$ for the decay of $^{71}$Ge back to $^{71}$Ga), and $F$ is the instantaneous flux or capture rate in SNU. We use the symbol $Q$ rather than $N$ to emphasize that the expected number is not an integer. The decay process included in Equation (1) must be taken into account in analyses of possible time variation of the neutrino flux. (See, for instance, Sturrock, Walther, and Wheatland 1997.)

These atoms will be extracted from the tank and transferred to an analysis chamber. However, the extraction process will not be perfect, and some of the capture-produced atoms will decay between extraction from the tank and the commencement of the analysis procedure. We assume that, when analysis begins, the expected number of capture-produced atoms in the analysis chamber, $Q_A$, is only K (a transfer coefficient) times the number produced in the tank, so that

$$Q_A = 10^{-36} K N_T \int_{t_S}^{t_E} dt\, F(t) e^{-\lambda(t_E - t)} . \tag{2}$$

### 3. Analysis of a Single Run

Events detected in the analysis chamber are of two types: they are due either to a background, which we here assume to be constant (this assumption can be relaxed), or to the decay of the target (capture-produced) atoms. We assume that the probability of an event due to the background occurring in a time interval $dt$ is $b\,dt$. We consider, for simplicity, just one type of event (typically an Auger event) from the target atoms. At the beginning of the analysis interval, the probability that a target atom will decay in the time interval $dt$ is $f\,dt$, where

$$f = \lambda Q_A . \tag{3}$$

We assume that the analysis chamber is activated for the time interval 0 *to T*, and that $N$ decay events are recorded at times $t_n$, where $n = 1,\ldots, N$. Our goal is to obtain an expression for the final probability distribution function $P_f(f \mid D)$, such that the probability that f is in the range f to $f + df$ is given by

$$P(f\ to\ f + df \mid D) = P_f(f \mid D) df , \tag{4}$$

where D denotes all relevant experimental data, *including* the fact that both *b* and *f* are non-negative.

Following Cleveland (1983), we introduce short time intervals $dt_n$, such that event *n* is taken to have occurred between $t_n$ and $t_n + dt_n$. Then the probability of an event occurring in that time interval is given by

$$P(E \mid t_n, t_n + dt_n) = \left(b + fe^{-\lambda t_n}\right) dt_n . \tag{5}$$

The probability that no event occurs before $t_1$ is

$$P(NE \mid 0, t_1) = \exp\left[-\int_0^{t_1} dt \left(b + fe^{-\lambda t}\right)\right] , \tag{6}$$

i.e.

$$P(NE \mid 0, t_1) = \exp\left[-bt_1 - \lambda^{-1} f \left(1 - e^{-\lambda t_1}\right)\right] . \tag{7}$$

The probability that no event occurs between $t_n$ and $t_{n+1}$ is given by

$$P(NE \mid t_n, t_{n+1}) = \exp\left[-b(t_{n+1} - t_n) - \lambda^{-1} f \left(e^{-\lambda t_n} - e^{-\lambda t_{n+1}}\right)\right], \tag{8}$$

and the probability that no event occurs between $t_N$ and T is given by

$$P(NE \mid t_N, T) = \exp\left[-b(T - t_N) - \lambda^{-1} f \left(e^{-\lambda t_N} - e^{-\lambda T}\right)\right] . \tag{9}$$

On multiplying the probabilities of all events and all non-events, we arrive at the likelihood of the data as a function of the parameters *b* and *f*:

$$L(D \mid b, f) = \exp\left[-bT - \lambda^{-1} f \left(1 - e^{-\lambda T}\right)\right] \prod_{n=1}^{N} \left(b + fe^{-\lambda t_n}\right) dt_n . \tag{10}$$

This expression is equivalent to Equation (13) of Cleveland (1983), and may be regarded as a special case of the "Cash likelihood" (Cash 1979).

The "dataset" $D$ comprises the times of the events, $t_1,\ldots,t_N$, and the time intervals $dt_1,\ldots,dt_N$. Since the prescribed time intervals do not depend on the parameters $b$ and $f$, we may ignore them, and consider instead the function

$$\tilde{L}(D|b,f) = \exp\left[-bT - \lambda^{-1}f\left(1-e^{-\lambda T}\right)\right]\prod_{n=1}^{N}\left(b + fe^{-\lambda t_n}\right). \quad (11)$$

If the prior probability distributions for $b$ and $f$ are written as

$$P(b \text{ to } b + db) = P_b(b)db \quad (12)$$

and

$$P(f \text{ to } f + df) = P_f(f)df, \quad (13)$$

then, by Bayes theorem (see, for instance, Howson and Urbach 1989), the post-pdf for $f$ is given by

$$P_f(f|D,b) = \frac{\tilde{L}(D|b,f)}{\int \tilde{L}(D|b,f')P_f(f')df'}P_f(f), \quad (14)$$

since the prior pdf for $f$ is assumed not to depend upon $b$.

Finally, the post pdf for $f$ is found by what is known as the "marginalization of the nuisance parameter," in this case the background rate $b$:

$$P_f(f|D) = \int P_f(f|D,b)P_b(b)db. \quad (15)$$

Note that the pdf's $P_b(b)$ and $P_f(f)$ are zero for negative values of their arguments. *This guarantees that the post pdf for f will be zero for negative values of f.*

## 4. Analysis of Many Runs

We now wish to combine data from several runs, which we enumerate by $r$, $r = 1,...,R$. Since the runs provide independent information, the pdf that represents the result of combining information from all runs is given by

$$P_f(f|D_1,...,D_R) = \frac{[P_f(f)]^{-(R-1)} P_f(f|D_1)...P_f(f|D_R)}{\int df' [P_f(f')]^{-(R-1)} P_f(f'|D_1)...P_f(f'|D_R)}. \tag{16}$$

(See, for instance, Sturrock (1973).) If we choose a flat form for the prior pdf for $f$, this becomes

$$P_f(f|D_1,...,D_R) = \frac{P_f(f|D_1)...P_f(f|D_R)}{\int df' P_f(f'|D_1)...P_f(f'|D_R)}. \tag{17}$$

That is to say, we simply form the product of the pdf's for the individual runs, and then normalize the result to integrate to unity.

*One may notice that this Bayesian procedure comprises a "global" analysis without the necessity of considering the timing data for all runs as one huge calculation, as is required by the standard likelihood procedure.*

One may also notice that it is very simple to update the pdf when data for a new run becomes available. We see from (17) that

$$P_f(f|D_1,...,D_R) = \frac{P_f(f|D_1,...D_{R-1}) P_f(f|D_R)}{\int df' P_f(f'|D_1,...D_{R-1}) P_f(f'|D_R)}. \tag{18}$$

We simply multiply the pdf that represents the results for the first $R-1$ runs by the pdf for the $R$'th run, and then normalize.

We see from Equations (2) and (3) that, once we have obtained an estimate of f, we may derive from it an estimate in SNU of the "flux" or "specific capture rate," assumed to be constant, from

$$F_{est} = 10^{36} K^{-1} N_T^{-1} \left[1 - e^{-\lambda(t_E - t_S)}\right]^{-1} f. \tag{19}$$

## 5. Incorporating Supplementary Information

An advantage of the Bayesian approach is that one can often incorporate supplementary information in one's calculation by suitable modification of the prior pdf's. This makes it possible to incorporate in our current calculation information that in the likelihood approach leads to negative flux estimates, which are of course unphysical and which one would like to avoid.

Some of the information available to the experimenters concerns the production of active atoms ($^{37}$Ar or $^{71}$Ge) by processes other than solar neutrinos. The experimenters sometimes take account of this supplementary information by subtracting the known non-solar contribution from the estimates of the total count rate obtained by the likelihood procedure. (See, for instance, Anselmann 1992), but this leads to negative estimates of the solar neutrino flux. We now show that the Bayesian approach offers an alternative procedure that avoids this problem.

Suppose that we are told that part of the production of active atoms is due to known non-solar sources, and that this production rate is the same as would be produced by a count rate $f_{ns}$ in the analysis chamber. Then the count rate f is given by

$$f = f_{ns} + f_{sol}, \tag{20}$$

where $f_{sol}$ is the actual solar contribution. Since $f_{sol} \geq 0$, it is sensible to require that the prior pdf for f should satisfy the requirement that

$$P_f(f) = 0 \text{ for } f < f_{ns}. \tag{21}$$

For a reasonable upper limit $f_{max}$ on $f$, we could for instance adopt a prior pdf such as

$$P_f(f) = 0 \text{ for } f < f_{ns},$$
$$P_f(f) = \frac{1}{f_{max} - f_{ns}} \text{ for } f_{ns} < f < f_{max}, \tag{22}$$
$$P_f(f) = 0 \text{ for } f > f_{max}.$$

The post-pdf for f, $P_{post,f}(f \mid D)$, will satisfy the same condition [Equation (21)] as the prior pdf:

$$P_{post,f}(f \mid D) = 0 \text{ for } f < f_{ns}. \tag{23}$$

From the post-pdf for f, we may obtain the post-pdf for the solar contribution to the count rate $f_{sol}$ from

$$P_{sol,f}(f_{sol} \mid D) = P_{post,f}(f_{ns} + f_{sol} \mid D). \tag{24}$$

In view of Equation (23), it follows that

$$P_{sol,f}(f_{sol} \mid D) = 0 \text{ for } f_{sol} < 0. \tag{25}$$

Hence negative values of the estimated solar neutrino flux are disallowed.

The actual situation concerning non-solar radioactive atoms is a little more complicated than the situation presented above. The non-solar contribution is not known exactly, but it can be represented by a pdf. One may replace this pdf by a small discrete set of values, each value being weighted according to the pdf. One may then carry out the procedure outlined earlier in this section for each of these values, giving the appropriate weight to the resulting post pdf for the flux, and finally combining the weighted post pdf's.

We may also note that the main reason that one obtains negative flux estimates for some runs is that the background rate sometimes increases during the analysis process (Pandola, 2007). One could allow for this effect by allowing by replacing Equation (12) by

$$P(b \text{ to } b + db) = db \iint db_0 db_1 P_{b_0}(b_0) P_{b_1}(b_1) \delta(b - b_0 - b_1 t), \tag{26}$$

and then marginalizing over both parameters $b_0$ and $b_1$. Alternatively, if a second analysis chamber is in operation, identical to the first except that no capture-produced atoms have been introduced, information concerning the background, expressed in terms of a pdf, could be used in the analysis of the neutrino-produced counts.

## 6. Discussion

The Bayesian approach appears to have advantages over the usual likelihood approach. In particular, it seems to be free from the four objections that were specified in Section 1. As a note of caution, it should be pointed out that estimates obtained in this way are not unbiased. For instance, if the solar neutrino flux is actually zero, the procedure we have outlined will necessarily lead to a nonzero estimate of the flux. This is not an uncommon situation in such analyses. (See, for instance, Bernado and Smith, 2004.)

We plan to examine the convenience and accuracy of our proposed procedure by Monte Carlo calculations, similar to those carried out by Cleveland (1983), and to compare the results with those obtained by the likelihood procedure. This work will be presented in a later article.

## Acknowledgements

We wish to thank Bruce Cleveland, Luciano Pandola and Jeffrey Scargle for helpful comments on this work, which was supported by NSF Grant AST-0607572.

**Appendix. Maximum-Likelihood Flux Estimates**

As we interpret the GNO data, we may construct a probability distribution function (pdf) for the flux measurement $F$ for each run (runs enumerated by $r$, $r = 1,...,R$) of the following form:

$$P_{F,r}(F)dF = \frac{(2/\pi)^{1/2}}{(\sigma_{l,r} + \sigma_{u,r})} \exp\left[-\frac{1}{2}\frac{(F - g_r)^2}{\sigma_{l,r}^2}\right] dF \quad \text{for } F < g_r$$

$$P_{F,r}(F)dF = \frac{(2/\pi)^{1/2}}{(\sigma_{l,r} + \sigma_{u,r})} \exp\left[-\frac{1}{2}\frac{(F - g_r)^2}{\sigma_{u,r}^2}\right] dF \quad \text{for } F > g_r$$

(A1)

where $g_r$ is the flux at which the pdf is a maximum and $\sigma_{u,r}$ and $\sigma_{l,r}$ are the upper and lower error estimates, respectively, for run r.

Ignoring the terms $dF$, which are independent of the variable $F$, we may form a log-likelihood function that combines data from all runs:

$$L(F) = \sum_r \ln[P_{f,r}(F)] .$$

(A2)

We find that this function has its peak value at 44.8 SNU, and that the curvature at the peak corresponds to an error estimate of 4.9 SNU.